\documentclass[twocolumn,10pt]{article}
\usepackage[utf8]{inputenc}
\usepackage[dvips]{graphicx}
\usepackage{amsthm}
\usepackage{amsmath}
\usepackage{amssymb}
\usepackage{natbib}

\newcommand{\ud}{\mathrm{d}}

\title{Example of an inhomogeneous cosmological model in the context of backreaction}
\author{Szymon Sikora, Krzysztof Głód \\
\scriptsize{\textit{Astronomical Observatory, Jagiellonian University, Orla 171, 30244 Krakow, Poland}} }

\begin{document}

\twocolumn[
\begin{@twocolumnfalse}
\maketitle
\begin{abstract}
In this article, we present an example of an inhomogeneous cosmological model, which is inspired by the linear perturbation theory. The metric of this model can be described as the \mbox{Einstein}-\mbox{de Sitter} background with a periodically distributed dust overdensities. The model construction enables application of the Green-Wald averaging scheme and the Buchert averaging technique simultaneously. We compare the angular diameter distance function of the considered model to the angular diameter distances corresponding to the \emph{average} space-times given by the Green-Wald and the Buchert frameworks respectively.
\end{abstract}
\vspace{1cm}
\end{@twocolumnfalse}
]

\small
\section{Introduction}
The impact of the matter inhomogeneities on the evolution of the Universe on a large scales is still a subject of debate. Among many different approaches to averaging in cosmology, it is worth to mention the Green-Wald scheme \cite{2011PhRvD..83h4020G} and the Buchert averaging method \cite{2000GReGr..32..105B}. In recent years there is a very interesting dispute between Green, Wald and their opponents \cite{2015CQGra..32u5021B,2015arXiv150606452G,2015arXiv151202947O,2016CQGra..33l5027G}. Green and Wald presented in \cite{2011PhRvD..83h4020G} a mathematically strict framework, which enables them to prove some important theorems. They show that the effective energy-momentum tensor, which appears in the Einstein equations due to their averaging procedure, is traceless and therefore cannot mimic the cosmological constant. This result contradicts alternative explanations of the accelerated expansion of the universe, like e.g. \cite{2013JCAP...10..043R}. On the other hand \cite{2015CQGra..32u5021B} argue that it is not clear whether the strict assumptions of Green and Wald apply to the real Universe, and the backreaction problem in cosmology is still open.  

There are several approaches to the presented problem. First, there is a growing interest on the numerical simulations concerning nonhomogeneous universe, \mbox{e.g. \cite{2016PhRvL.116y1302B,2016PhRvL.116y1301G,2016PhRvD..93l4059M}}. These papers show that the matter inhomogeneities can produce some important deviations from the homogeneous models on sub-Hubble scales. Another thing is to study the propagation of light in the inhomogeneous space-times (e.g. \cite{2015JCAP...11..022F,2016arXiv160804403G,2011PhRvD..84d4011S}) . Finally, it is important to analyze some model space-times, to which the averaging procedures can be applied. It is believed that the metric of our Universe could be very complicated, but in the practical usage of the averaging techniques usually the knowledge of the true Universe metric is not necessary. In the viewpoint of the Green and Wald framework, it is sufficient that there exist a family of metrics which describe the true Universe and fulfills the specified assumptions. The properties of the effective energy-momentum tensor follow from the presented theorems and they are true for any family of metrics satisfying the model assumptions. In the case of the Buchert approach, one can solve the Buchert equations with some closure conditions, and find the effective scale factor without knowledge of the original metric. However, to clarify the issue of a backreaction it is worth to study simple models, in which the space-time metric is given explicitly. Examples of such a nontrivial space-times, to which the Green-Wald scheme apply were given in \cite{2014PhRvD..89d4033S,2016PhRvD..94b4059S}. 

The aim of this paper is to present a very simple inhomogeneous space-time with a metric given explicitly, to which the Green-Wald and Buchert schemes both apply and for which the Friedmann-Lema\^{i}tre-Robertson-Walker (FLRW) space-time is expected as the average.  

This paper is organized as follows. In Section 2, we present the metric of our model and its basic properties. In Section 3, we show how the Green-Wald and Buchert procedures can be applied to the space-time under consideration. Section 4 is dedicated to the comparison between the angular diameter distances calculated in a presented space-time and derived in the Green-Wald and Buchert average space-times respectively.

\section{The model}
Our model space-time is inspired by the perturbation theory. We assume that the metric can be written as\footnote{We use the convention in which greek letters label the indices which cover the range $\{0,1,2,3\}$, while the latin letters describe the space-like indices $\{1,2,3\}$.}:
\begin{equation}\label{ref:metric1}
g_{\mu\:\!\nu}=g^{(0)}_{\mu\:\!\nu}+\lambda\,h_{\mu\:\!\nu}\,,
\end{equation}
where $g^{(0)}$ represents the \mbox{Einstein}-\mbox{de Sitter} background metric. To be able to perform the Buchert averaging we introduce the coordinates $\{t,x,y,z\}$ in which the background metric reads: $g^{(0)}_{\mu\:\!\nu}=\mathrm{diag}(-1,a^2,a^2,a^2)$, with a scale factor $a(t)=\mathcal{C}\,t^{2/3}$ and $\mathcal{C}$ being a constant. We will use the natural units $c=1$ and $G=1$. The remaining tensor $h_{\mu\:\!\nu}$ we define with the help of the two scalar functions $C(t,x,y,z)$ and $D(t,x,y,z)$ as follows\footnote{We adopt the convention of the partial and covariant derivatives where $f_{,x}\equiv\partial_x f$ and $f_{;x}\equiv\nabla_x f$ }:
\begin{equation}
h_{0\:\!0}=0\,, \quad h_{i\:\!0}=0\,,
\end{equation}
\begin{equation}
h_{i\:\!j}=a(t)^2\,\left(C_{,i\:\!j}-\frac{1}{3}\delta_{i\:\!j}(C_{,xx}+C_{,yy}+C_{,zz})+\delta_{i\:\! j}\,D \right)\,. \nonumber
\end{equation}
We put the functions $C$ and $D$ as:
\begin{equation}
C(t,x,y,z)=-\frac{\mathcal{C}^3}{81\,t}\left(f(x)+f(y)+f(z)\right)\,,
\end{equation}
\begin{equation}
D(t,x,y,z)=-\frac{\mathcal{C}^3}{243\,t}\left(\frac{\ud^2 f(x)}{\ud x^2}+\frac{\ud^2 f(y)}{\ud y^2}+\frac{\ud^2 f(z)}{\ud z^2} \right)\,,\nonumber
\end{equation}
with:
\begin{equation}\label{ref:metric4}
f(w)=\frac{w^2}{16}+\frac{1}{32\,B^2}\,\cos(2\,B\,w)\,.
\end{equation}
The metric defined this way depends on the two free parameters $\lambda$ and $B$.

For a metric (\ref{ref:metric1}), the Einstein tensor can be expanded in a Taylor series around $\lambda=0$:
\begin{equation}
G_{\mu\:\!\nu}=G^{(0)}_{\mu\:\!\nu}+\lambda\,G^{(1)}_{\mu\:\!\nu}+\lambda^2\,G^{(2)}_{\mu\:\!\nu}+\dots\,.
\end{equation}
When one takes the similar decomposition of the energy-momentum tensor:
\begin{equation}
T_{\mu\:\!\nu}=T^{(0)}_{\mu\:\!\nu}+\lambda\,T^{(1)}_{\mu\:\!\nu}+\lambda^2\,T^{(2)}_{\mu\:\!\nu}+\dots\,,
\end{equation}
then it is easy to distinguish the background energy-momentum tensor $T^{(0)}_{\mu\:\!\nu}=G^{(0)}_{\mu\:\!\nu}/8\pi$. It has a form of the dust $T^{(0)}_{\mu\:\!\nu}=\rho^{(0)}\,U_\mu\,U_\nu$, with a four-velocity of the observer comoving with matter $U^\mu=(1,0,0,0)$ and the background density $\rho^{(0)}=(4/3)\,t^{-2}$. The ansatz (\ref{ref:metric1}-\ref{ref:metric4}) is proposed in such a way that the first order energy-momentum tensor $T^{(1)}_{\mu\:\!\nu}=G^{(1)}_{\mu\:\!\nu}/8\pi$ has also the form of the dust: $T^{(1)}_{\mu\:\!\nu}=\rho^{(1)}\,U_\mu\,U_\nu$, with the density given by:
\begin{equation}
\rho^{(1)}=\frac{\mathcal{C}^3}{3888\,\pi\,t^3}\,\left(\sin^2(B\,x)+\sin^2(B\,y)+\sin^2(B\,z) \right)\,.
\end{equation}
Note, that we assumed that the observer four-velocity $U^\mu$ is not perturbed (it does not depend on $\lambda$ and it is tangent to the time-like geodesic since $\Gamma^\mu_{0\:\!0}=0$). 
\begin{figure}
   \centering
      \includegraphics[width=0.4\textwidth]{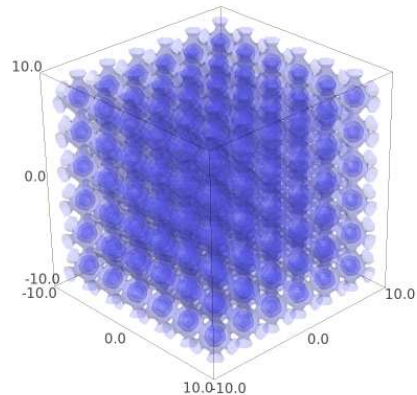}
      \caption{\label{fig:isodensity} \scriptsize{Isodensity surfaces of the function $\sin^2(x)+\sin^2(y)+\sin^2(z)$, which is a distribution of the first order density perturbation $\rho^{(1)}$.}}
\end{figure}
The exemplary isodensity surfaces of $\rho^{(1)}$ are plotted in \mbox{Figure \ref{fig:isodensity}}. They form a periodic, cubic lattice. The parameter $B$ controls the size of the elementary cell, which is related directly to the size of the overdensities. The parameter $\lambda$ gives the amplitude of the overdensities. There is an important constraint concerning $\lambda$. We want to guarantee that the second and higher order energy-momentum tensor components are small in comparison to the energy density up to the first order:
\begin{equation}\label{ref:condition_second_order}
\rho^{(0)}+\lambda\,\rho^{(1)}\,\gg\,|\lambda^2\,T^{(2)}\:\!{}^\mu\:\!{}_\nu+\dots\,|\,.
\end{equation}
In the order $k\geq 2$, the tensor $T^{(k)}_{\mu\:\!\nu}=G^{(k)}_{\mu\:\!\nu}/8\pi$ has a more complicated form, but when the condition (\ref{ref:condition_second_order}) holds, the proposed metric can be thought as a metric approximately well describing some dust inhomogeneous cosmological model, to which the Buchert averaging \cite{2000GReGr..32..105B} can be applied. In Section \ref{sec:ModelParameters}, we specify more precisely when this condition is satisfied.

Before going further let us justify the choice of the density perturbation $\rho^{(1)}$. The periodically distributed overdensities correspond to the discrete counterpart of the FLRW symmetry at the hypersurface of a constant time. Moreover, at the scales much larger than the size of the elementary cell the distribution of mass is uniform in common sense. Such a periodically distributed matter have been studied in a case of the black hole lattices \cite{2014CQGra..31h5002K} and within numerical simulations with periodic boundary conditions \cite{2016PhRvL.116y1302B}.

We want to notice also that we treat our space-time only as a toy model. It has no ambition to describe the true Universe (e.g. the density contrast $\rho^{(1)}/\rho^{(0)}$ falls off like $t^{-1}$, so the model does not describe the formation of structures and it is useless for describing the early universe). Nevertheless, the model is very useful to see the averaging schemes in action and could be generalized in the future.

\section{The average space-times}
\subsection{The Green-Wald scheme}
We use the formula (\ref{ref:metric1}) as a definition of the Green-Wald's family of metrics $g_{\mu\:\!\nu}(\lambda)$. The conditions \emph{(i)-(iv)} which this family of metrics should satisfy are listed in \cite{2011PhRvD..83h4020G}. \emph{(i)} For every positive $\lambda$ the Einstein equations hold, and the background metric $g^{(0)}$ defined by the limit $\lambda\to 0$ is the \mbox{Einstein}-\mbox{de Sitter} one. The weak energy condition is satisfied up to the first order $T^{(0)}_{\mu\:\!\nu}+\lambda\,T^{(1)}_{\mu\:\!\nu}$ since this energy-momentum tensor represents a dust with a positive density.  The second order energy-momentum tensor $T^{(2)}_{\mu\:\!\nu}$ is diagonal, and its elements are always positive. This shows that the weak energy condition is fulfilled up to the second order too. In the higher orders, the energy-momentum components become more complicated
and it is not easy to prove that the weak energy condition is satisfied in general. However, according to the assumption (\ref{ref:condition_second_order}), the higher order energy-momentum tensor components are irrelevant for our considerations. \emph{(ii)} The metric perturbation $g_{\mu\:\!\nu}(\lambda)-g^{(0)}_{\mu\:\!\nu}$ has the form $\lambda\,h_{\mu\:\!\nu}$. When we restrict ourselves to $t>t'$, where $t'$ can be arbitrary small, then $h_{\mu\:\!\nu}$ is bounded. In practice we are interested in the late times only, so the time $t'$ can be taken for example as $t'\approx 3\,\mathrm{Gyr}$. \emph{(iii-iv)} The covariant derivative of $g_{\mu\:\!\nu}(\lambda)-g^{(0)}_{\mu\:\!\nu}$ is proportional to $\lambda$ and in the limit $\lambda\to 0$ tends to zero, so the components of the tensor $\mu_{\alpha\:\!\beta\:\!\gamma\:\!\delta\:\!\epsilon\:\!\phi}$ are equal to zero. This shows that the conditions \emph{(i)-(iv)} are satisfied. 

According to \cite{2011PhRvD..83h4020G}, in effect of the properly defined weak limit $\mathrm{\hbox{w-lim}}\,\lambda\to0$, in the Einstein equations concerning $g^{(0)}$ appears the additional \emph{effective} energy-momentum tensor $t^{(0)}_{\alpha\:\!\beta}$, which represents the impact of the inhomogeneities on the global dynamics. Because components of $\mu_{\alpha\:\!\beta\:\!\gamma\:\!\delta\:\!\epsilon\:\!\phi}$ are equal to zero, from the Equation 16 in \cite{2011PhRvD..83h4020G} the effective energy-momentum tensor $t^{(0)}_{\alpha\:\!\beta}$ in our case is equal to zero also. There is no backreaction (the inhomogeneities does not influence the global dynamics of the space-time). Although this example is trivial in the viewpoint of the Green-Wald scheme, it is interesting to compare the \mbox{Einstein}-\mbox{de Sitter} metric (which can be thought as the average space-time here) with the predictions of the Buchert method. 

\subsection{The Buchert scheme}
The main idea behind the Buchert approach concerns the fact, that having the three-dimensional metric $g^{(3)}_{i\:\!j}$ induced on the hypersurface of a constant time $t$ one can define the spatial volume of some domain $\mathcal{D}(t)$:
\begin{equation}
V_\mathcal{D}(t)=\int_{\mathcal{D}(t)}\,\sqrt{\mathrm{det}\,g^{(3)}}\,\ud^3 X\,,
\end{equation}
and a spatial average of any scalar function $\Psi$:
\begin{equation}
\langle \Psi \rangle_\mathcal{D}=\frac{1}{V_\mathcal{D}}\,\int_{\mathcal{D}(t)}\,\sqrt{\mathrm{det}\,g^{(3)}}\,\Psi(t,X^k)\,\ud^3 X\,,
\end{equation}
where the coordinate choice is $x^\mu=(t,X^k)$. Following \cite{2000GReGr..32..105B}, one may consider the universe filled with the dust. It is described by the energy-momentum tensor $T_{\mu\:\!\nu}=\rho\,U_\mu\,U_\nu$, where the four-velocity of the observer comoving with matter is $U^\mu=(1,0,0,0)$. By applying the above type of averaging to the scalar part of the Einstein equations (the Raychaudhuri equation and the Hamiltonian constraint) one can derive the two Friedmann-like equations:
\begin{equation}\label{ref:BuchertEqn1}
3\frac{\ddot{a}_\mathcal{D}}{a_\mathcal{D}}+4\,\pi\,G\,\langle\rho\rangle_\mathcal{D}=\mathcal{Q}_\mathcal{D}\,,
\end{equation}
\begin{equation}\label{ref:BuchertEqn2}
3\,\left(\frac{\dot{a}_\mathcal{D}}{a_\mathcal{D}} \right)^2-8\,\pi\,G\,\langle\rho\rangle_\mathcal{D}+\frac{1}{2}\langle\mathcal{R}\rangle_\mathcal{D}=-\frac{\mathcal{Q}_\mathcal{D}}{2}\,.
\end{equation}
For simplicity, we put the cosmological constant $\Lambda=0$. In the above equations appears the \emph{effective scale factor}:
\begin{equation}
a_\mathcal{D}(t)=\left(\frac{V_\mathcal{D}(t)}{V_\mathcal{D}(t_0)} \right)^{1/3}\,,
\end{equation}
which is related to the actual volume of the domain normalized to the volume at the present time $t_0$. The $\mathcal{R}$ is a spatial Ricci scalar, while the term $\mathcal{Q}_\mathcal{D}$ represents the backreaction. In our case, when the metric is given, we will calculate $\mathcal{Q}_\mathcal{D}$ in the following steps. The tensor $P_{\mu\:\!\nu}=g_{\mu\:\!\nu}+U_\mu\,U_\nu$ projects onto the space orthogonal to $U_\mu$. First, we calculate the components of the extrinsic curvature tensor from: $K_{i\:\!j}=-P^\mu{}\,_i\,P^\nu{}_j\,U_{\mu;\nu}$. This enables us to obtain the expansion scalar $\theta=-K^i{}_i$, the shear tensor $\sigma_{i\:\!j}=-K_{i\:\!j}-\theta\,g^{(3)}_{i\:\!j}/3$ and the rate of shear $\sigma=\sigma^i{}_j\,\sigma^j{}_i/2$. Then we construct the two scalar invariants: $\mathrm{\textbf{I}}=\theta$ and $\mathrm{\textbf{II}}=\theta^2/3-\sigma^2$ and calculate their averages. The backreaction term is given by Equation 10c in \cite{2000GReGr..32..105B}:
\begin{equation}
\mathcal{Q}_\mathcal{D}=2\,\langle\mathrm{\textbf{II}}\rangle_\mathcal{D}-\frac{2}{3}\langle\mathrm{\textbf{I}}\rangle_\mathcal{D}^2
\end{equation}
For the purposes of this article, by the Buchert average space-time we mean the space-time described by the FLRW metric $\widetilde{g}_{\mu\:\!\nu}=(-1,a^2_\mathcal{D},a^2_\mathcal{D},a^2_\mathcal{D})$, where the effective scale factor $a_\mathcal{D}$ satisfies the Buchert equations (\ref{ref:BuchertEqn1},\ref{ref:BuchertEqn2}), and the domain $\mathcal{D}$ is the elementary cell (because of the periodicity, this is equivalent to averaging over the whole space).

\subsection{The specific values of the model parameters}\label{sec:ModelParameters}
To show some quantitative results we have to specify the model parameters. As we mentioned in the introduction we use the natural units $c=1$ and $G=1$. We set the megaparsec as a unit of length. For a \mbox{Hubble} constant $H_0=67.3\,\mathrm{km/s/Mpc}$ \cite{2014A&A...571A..16P}, the age of the \mbox{\mbox{Einstein}-\mbox{de Sitter}} universe $t_0=2/(3H_0)=9.67\,\mathrm{Gyr}$ in these units reads $t_0=2969.7\,\mathrm{Mpc}$. The usual convention of the scale factor scaling $a(t_0)=1$ sets then the value of the constant $\mathcal{C}=0.0048$. We fix the parameter $B=1$, so the elementary cell is a domain $\mathcal{D}$ defined by the inequalities $0\leq x<\pi$, $0\leq y<\pi$ and $0\leq z<\pi$, and the distance around $1.57\,\mathrm{Mpc}$ can be thought as a radius of the overdensity region at $t_0$. This is a typical scale for a galaxy cluster size. It is convenient to measure the density in the units of the critical density $\rho_{cr}=3\,H_0^2/(8\pi)$. The background density at $t_0$ in this units is $\Omega^{(0)}\equiv\rho^{(0)}/\rho_{cr}=1.0$. We fix the 
amplitude of the overdensities $\lambda$ by demanding that on the maximum $x=\pi/2$, $y=\pi/2$, $z=\pi/2$ the density in the first order is the one tenth of the background density in critical units $\Omega^{(1)}\equiv\lambda\,\rho^{(1)}/\rho_{cr}=0.1$. We choose this particular value because it is smaller than the background density and the average $\langle \Omega^{(1)}\rangle_\mathcal{D}=0.0497$ is not negligible. One can compare this number with the estimation of the total amount of the baryonic mass, derived from the primordial nucleosynthesis \cite{2001ApJ...552L...1B}. 

\begin{figure}[h]
   \centering
      \includegraphics[width=0.45\textwidth]{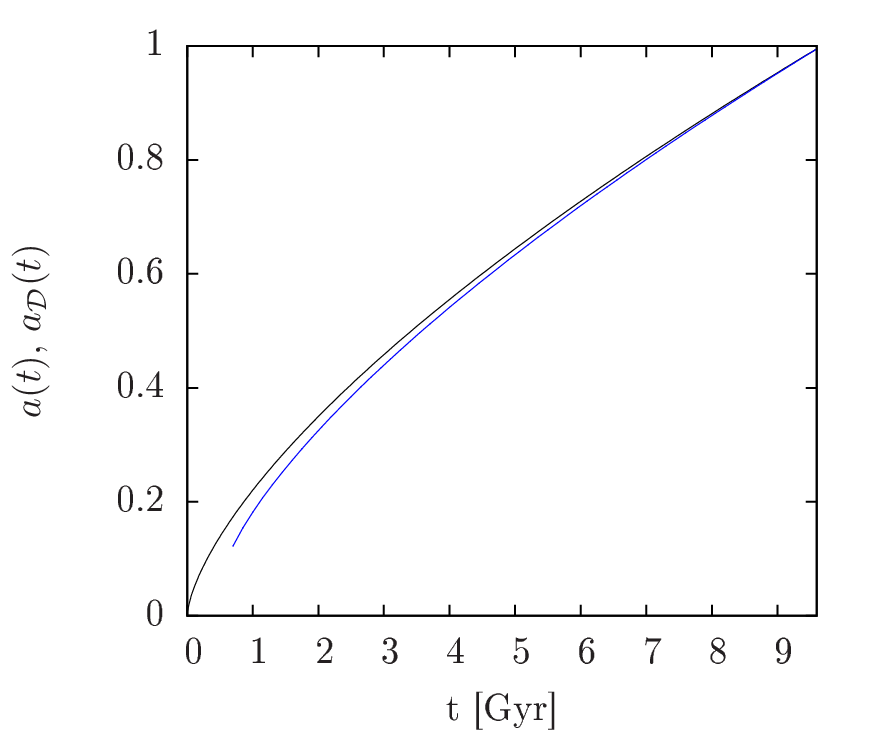}
      \caption{\label{fig:scale_factor} \scriptsize{Comparison between the effective scale factor $a_\mathcal{D}$ (\emph{blue}) and the background scale factor $a(t)$ (\emph{black}), for a model with parameters given in the text.}}
\end{figure}

Before we examine whether condition (\ref{ref:condition_second_order}) is satisfied for the choice $\Omega^{(1)}=0.1$, we show on the Figure \ref{fig:scale_factor} the bahaviour of the effective scale factor $a_\mathcal{D}$ compared with the scale factor $a(t)$.
In calculation of $a_\mathcal{D}$ we get the determinant $\mathrm{det}\,g^{(3)}$ directly from the metric (\ref{ref:metric1}) with parameters specified above and perform the numerical integration over the elementary cell as a domain $\mathcal{D}$. This procedure is somehow different from the customary approach, because instead of solving the Buchert equations (\ref{ref:BuchertEqn1},\ref{ref:BuchertEqn2}) in the derivation of $a_\mathcal{D}$, we obtain it directly from the proposed metric. For the effective scale factor $a_\mathcal{D}$ obtained this way we will verify the Buchert equations in the next paragraph. Figure \ref{fig:scale_factor} shows that the effective scale factor is slighlty lower than its background counterpart. For the times $t<0.7\,\mathrm{Gyr}$ the determinant $\mathrm{det}\,g^{(3)}$ becomes negative in some points. For the considered foliation of space-time, the $a_\mathcal{D}$ cannot be properly defined there. In this work, we won't analyze the structure of the initial singularity in the more details.

\begin{figure}[h]
   \centering
      \includegraphics[width=0.45\textwidth]{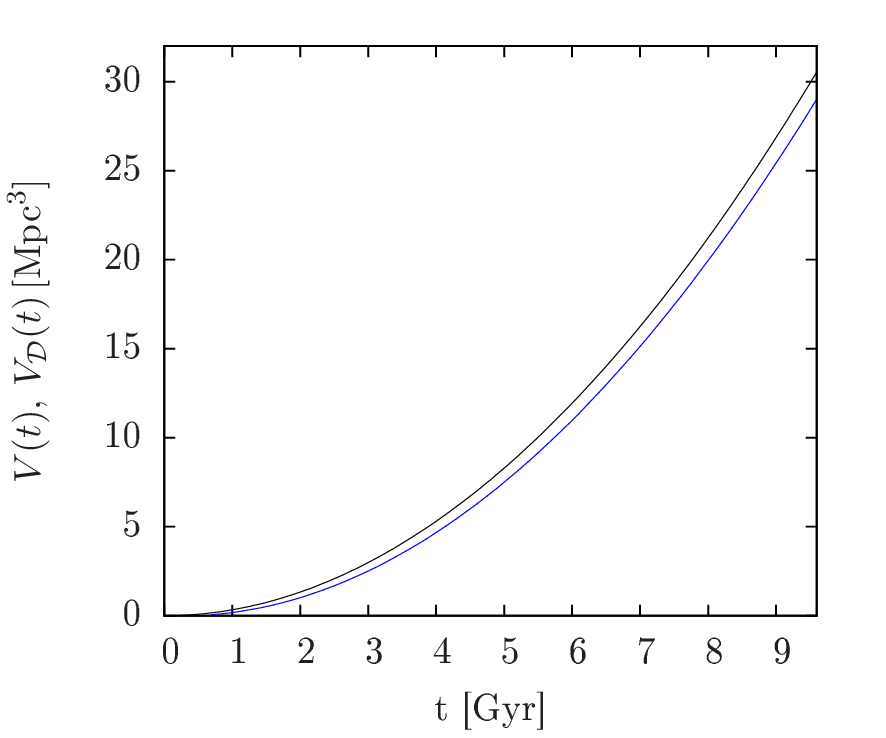}
      \caption{\label{fig:volumes} \scriptsize{Comparison between the volume of the elementary cell $x\in[0,\pi)$, $y\in[0,\pi)$, $z\in[0,\pi)$, calulated for the metric with parameters described in the text $V_\mathcal{D}$ (\emph{blue}), with the volume of the elementary cell in the background metric $V=\pi^3\,a^3(t)$ (\emph{black}).}}
\end{figure}

It is interesting also to see the comparison between the volume of the elementary cell in the considered model $V_\mathcal{D}$ and \mbox{Einstein}-\mbox{de Sitter} background $V$, which is plotted on the Figure \ref{fig:volumes}. The difference $V_\mathcal{D}-V$ at the time $t_0$ is $1.54\,\mathrm{Mpc}^3$, what corresponds to the relative difference of about $5\%$. 

Now, lets focus on the condition (\ref{ref:condition_second_order}). At first we compare the components of the second order energy-momentum tensor expressed in the critical units $\Omega^{(2)}_{\mu\:\!\nu}\equiv \lambda^2\,T^{(2)}_{\mu\:\!\nu}/\rho_{cr}$ calculated on the maximum of the overdensity $x=\pi/2$, $y=\pi/2$, $z=\pi/2$, at the time $t_0$, with the first order density $\Omega^{(1)}=0.1$ and the background density $\Omega^{(0)}=1.0$ there. The component corresponding to the density in the second order has the value $\Omega^{(2)}_{0\:\!0}=0.009$ and it is about ten times smaller than $\Omega^{(1)}$. 
The spatial distribution of $\lambda^2\,T^{(2)}_{0\:\!0}$ does not change the overall picture of the isodensity surfaces depicted on Figure \ref{fig:isodensity}. The pressure-like terms $\Omega^{(2)}\:\!{}^i\:\!{}_i$ are not larger than $0.001$, which is less than $0.1\%$ of the energy density $(\rho^{(0)}+\lambda\,\rho^{(1)})/\rho_{cr}$. The nondiagonal terms $\Omega^{(2)}_{i\:\!k}$, where $i\neq k$, are strictly equal to zero. This estimation shows that the condition (\ref{ref:condition_second_order}) is valid at the time $t_0$. 
\begin{figure}[h]
   \centering
      \includegraphics[width=0.45\textwidth]{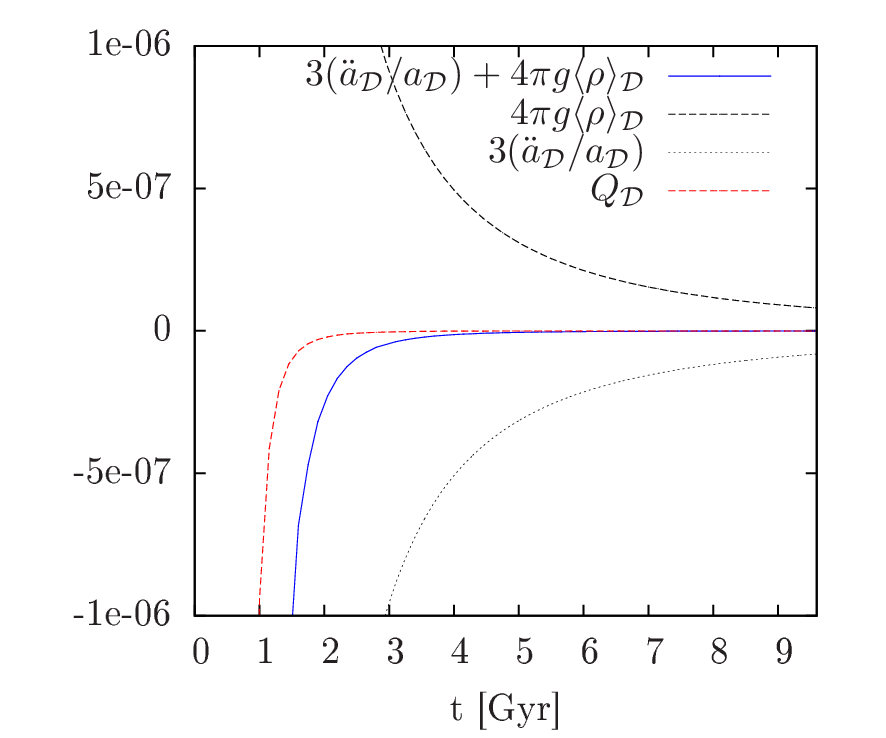}
      \caption{\label{fig:Buchert1} \scriptsize{The components of the first Buchert equation as a functions of time $t$. The left hand side of the equation is plotted by \emph{blue}, while the right hand side is plotted by \emph{red} curve.}}
\end{figure}
\begin{figure}[h]
   \centering
      \includegraphics[width=0.45\textwidth]{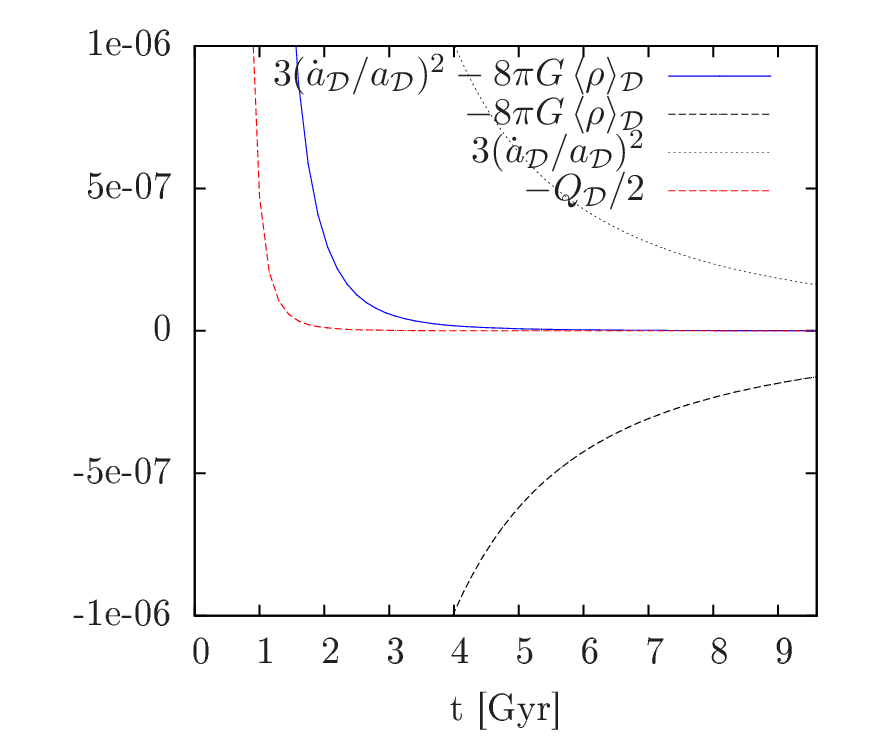}
      \caption{\label{fig:Buchert2} \scriptsize{The components of the second Buchert equation as a functions of time $t$. The left hand side of the equation is plotted by \emph{blue}, while the right hand side is plotted by \emph{red} curve.}}
\end{figure}

To check whether this condition is fulfilled for the earlier times $t<t_0$ we verify the Buchert equations. In Figures \ref{fig:Buchert1} and \ref{fig:Buchert2}, we show the components of the first Buchert equation (\ref{ref:BuchertEqn1}) and the second Buchert equation (\ref{ref:BuchertEqn2}) respectively, as a functions of the time $t$. The effective scale factor $a_\mathcal{D}$ and the backreaction term $\mathcal{Q}_\mathcal{D}$ were derived directly from the model metric, the density $\rho$ is taken up to the first order $\rho=\rho^{(0)}+\lambda\,\rho^{(1)}$, while the averages were calculated by means of the numerical integration over the elementary cell as a domain $\mathcal{D}$. The average spatial Ricci scalar $\langle\mathcal{R}\rangle_\mathcal{D}$ is $3.2\times10^{-18}$ at $t_0$, and about $1.7\times10^{-16}$ at $t=3\,\mathrm{Gyr}$, so concerning the second equation, it is negligible. At both figures, the left-hand side of the equation is plotted by the blue curve, while the right-hand side is drawn by the red curve. One can easly see that the Buchert equations are fulfilled at the late times, but they are not satisfied for the early times. This is because for a small $t$ the assumption (\ref{ref:condition_second_order}) is not valid. The higher order energy-momentum tensors become important there, and they have the pressure-like terms, whereas the Buchert equations (\ref{ref:BuchertEqn1},\ref{ref:BuchertEqn2}) were derived assuming the universe filled with the dust. On the other hand, at late times the components of the higher order energy-momentum tensors other than the density contribution $T^{(k)}_{0\:\!0}$ are negligible. The validity of the equations (\ref{ref:BuchertEqn1},\ref{ref:BuchertEqn2}) proves that the proposed metric (\ref{ref:metric1}-\ref{ref:metric4}) very well approximates the nonhomogeneous universe filled with the dust, if only the time $t$ is large enough. There is no sharp distinction between these two regimes, but the presented figures suggest us that from the time around $t'\approx 3\,\mathrm{Gyr}$ up to the age of the \mbox{Einstein}-\mbox{de Sitter} universe $t_0$ the condition (\ref{ref:condition_second_order}) is satisfied and the model describes the dust universe well. At the time $t=3\,\mathrm{Gyr}$ the pressure-like terms $\Omega^{(2)}\:\!{}^i\:\!{}_i$ are around $2\%$ of the first order density $(\rho^{(0)}+\lambda\,\rho^{(1)})/\rho_{cr}$ at that time. The following section is dedicated to some observables in the considered cosmological model. Based on these results one can see that the time $t=3\,\mathrm{Gyr}$ corresponds to the redshift $z=1.2$. It is reasonable then to compare the observables calculated in the proposed metric with the predictions obtained within the Buchert average space-time and the \mbox{Einstein}-\mbox{de Sitter} background, up to the redshift $z=1.2$.

\section{The observables}
\subsection{The redshift}\label{sec:redshift}
To calculate the redshift we produced numerically the family of one hundred null geodesics in the following way. For each geodesic $x^\mu(\lambda)$, we solve with the help of the fourth order Runge-Kutta method, the geodesic equation:
\begin{equation}
\frac{\ud k^\mu}{\ud\lambda}=-\Gamma^\mu_{\alpha\:\!\beta}\,k^\alpha\,k^\beta\,,
\end{equation}
and the equation defining the wave vector as a vector tangent to the geodesic:
\begin{equation}
\frac{\ud x^\mu}{\ud\lambda}=k^\mu\,.
\end{equation}
The initial conditions at $\lambda=0$ are taken such that the position of the observer is $x^\mu=(t_0,0,0,0)$, the timelike component of the wavevector is $k^0=-1$ (so the geodesic is past oriented), and the direction of the three-vector $k^i$ is generated randomly with the probability distribution uniform on the unit sphere (the constraint $k^i\,k_{\:\!i}=1$ guarantees that $k^\mu$ is a null vector at $\lambda=0$). During the numerical integration the condition $k^\mu\,k_\mu=0$ for each $\lambda>0$ is adopted as a test of the numerical convergence.

\begin{figure}[h]
   \centering
      \includegraphics[width=0.45\textwidth]{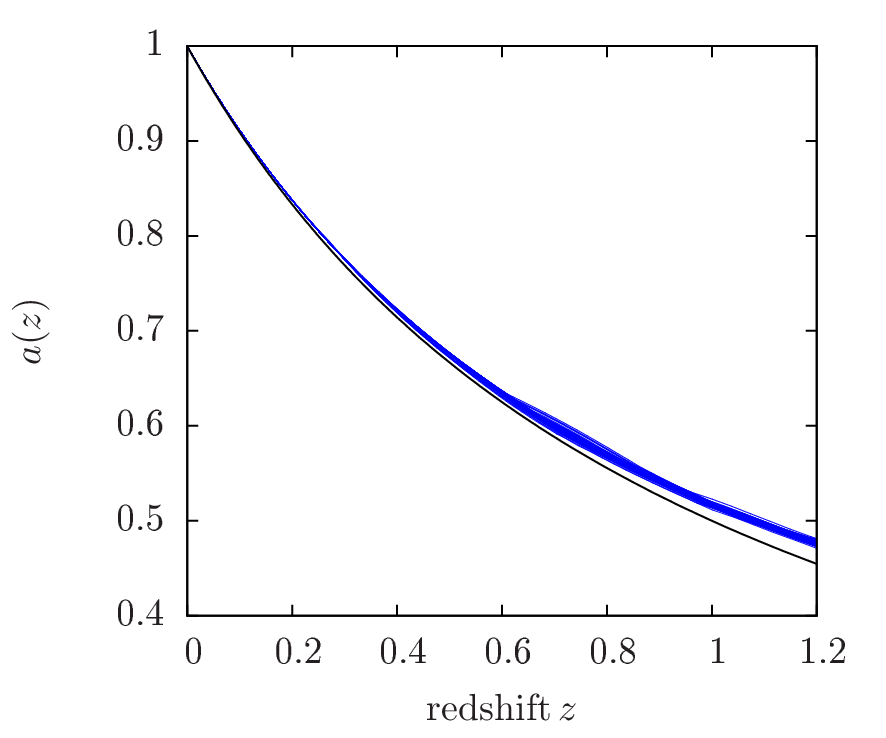}
      \caption{\label{fig:redshift} \scriptsize{The relation between the scale factor and the redshift. In \emph{blue} we plotted the family of the one hundred $a(z)$ curves, each of them corresponding to the particular geodesic. This result is compared with the standard relation $a(z)=1/(1+z)$ for a background \mbox{Einstein}-\mbox{de Sitter} universe (\emph{black}).}}
\end{figure}

Once we obtained the geodesic $x^\mu(\lambda)$, we may calculate the redshift. The observer four-velocity is $U^\mu=(1,0,0,0)$, so the observed frequency of the light at $\lambda=0$ is normalized to unity $\omega_{obs}=k^\mu\,U_\mu=1$. Suppose that the source of the light is located on the geodesic $x^\mu(\lambda)$ and its four-velocity is $U^\mu=(1,0,0,0)$ (so it is comoving with the matter also). Then the frequency of the emitted light is $\omega_{em}=k^\mu(\lambda)\,U_\mu=-k^0(\lambda)$ at the particular $\lambda$ (note, that because the geodesic is past oriented the $k_0$ is negative so that the frequency  $\omega$ is positive). The redshift is then $z=(\omega_{em}-\omega_{obs})/\omega_{obs}=-k^0(\lambda)-1$. This way the redshift can be obtained as a function of the affine parameter along the geodesic $z(\lambda)$. By taking into account the function $a(t)=\mathcal{C}\,t^{2/3}$, we may recover the relation between the redshift and the scale factor $a(z)$, where the redshift is taken along the particular geodesic. On the Figure \ref{fig:redshift} we plot this relation for a family of the one hundred geodesics (the blue curves), each of them generated numerically by the method given above. Because of a large number of geodesics considered, the Reader could have an impression that this picture represents one thick blue curve. The thickness of this curve is related to the spread of the redshifts caused by the fact that various geodesics pass through a different overdensity regions. The resulting $a(z)$ is slightly above the standard profile $a(z)=1/(1+z)$ corresponding to the background \mbox{Einstein}-\mbox{de Sitter} space-time. Note however, that in the presented model the function $a(t)$ has a different meaning than the scale factor which appears in the \mbox{Einstein}-\mbox{de Sitter} universe, because here $a(t)$ is only one of the three distinct metric functions, and its relation to physical distances is not so straightforward.       

\subsection{The angular diameter dinstance and the luminosity distance}
Lets consider the light beam, for which the central ray follows the geodesic $x^\mu(\lambda)$. The angular diameter distance $d_A=\sqrt{A_S/\Omega_O}$ by definition relates the area of the cross-section of the beam at the position of the light source $A_S$, with the solid angle of the beam $\Omega_O$, measured by the observer at $\lambda=0$. The luminosity distance\footnote{In literature it is often called uncorrected luminosity distance.} $d_L$ is defined as $d_L^2=L/(4\pi f)$, where $L$ is the source luminosity and $f$ is the flux measured by the observer at $\lambda=0$. With the help of the reciprocity theorem \cite{1992grle.book.....S} and by taking into account the correction factor due to the change of the energy of photons, one can derive the following relationship:
\begin{equation}\label{eqn:luminosity}
d_L=(1+z)^2\,d_A\,,
\end{equation}
where $z$ is a source redshift. The luminosity distance and the angular diameter distance are associated with one another, so it is sufficient to calculate one of these two quantities. In order to obtain $d_A$, one can follow the idea of Sachs \cite{1961RSPSA.264..309S}, which leads to the \emph{focusing equation} \cite{2004LRR.....7....9P}:
\begin{equation}\label{eqn:Focusing}
\frac{\ud^2}{\ud \lambda^2}\,d_A=-\left(\frac{1}{2}\,R_{\mu\:\!\nu}\,k^\mu\,k^\nu+|\sigma|^2 \right)\,d_A\,,
\end{equation}
where $\sigma$ is a complex shear satisfying:
\begin{equation}\label{eqn:shear1}
|\sigma|^2=\frac{1}{2}\,(\nabla_\mu\,k_\nu)(\nabla^\mu\,k^\nu)-\theta^2\,,
\end{equation}
and $\theta$ is expansion rate:
\begin{equation}\label{eqn:expansion1}
\theta=\frac{1}{2}\nabla_\mu\,k^\mu\,.
\end{equation}
Bacause we have found numerically only the central ray geodesic $x^\mu(\lambda)$, instead of using the formulas (\ref{eqn:shear1},\ref{eqn:expansion1}) we calculate the complex shear $\sigma=\sigma_1+\mathrm{i}\,\sigma_2$ on the geodesic $x^\mu(\lambda)$, from the Sachs scalar evolution equation \cite{2015JCAP...11..022F}:
\begin{equation}\label{eqn:Sachs_sigma}
\frac{\ud}{\ud\lambda}\sigma_1+2\,\sigma_1\,\theta=-\frac{1}{2}\,C_{\alpha\:\!\beta\:\!\gamma\:\!\delta}\,\left(s_1^\alpha\,k^\beta\,k^\gamma\,s_1^\delta+s_2^\alpha\,k^\beta\,k^\gamma\,s_2^\delta \right)\,,\nonumber
\end{equation}
\begin{equation}
\frac{\ud}{\ud\lambda}\sigma_2+2\,\sigma_2\,\theta=C_{\alpha\:\!\beta\:\!\gamma\:\!\delta}\,s_1^\alpha\,k^\beta\,k^\gamma\,s_2^\delta\,,
\end{equation}
where $C_{\alpha\:\!\beta\:\!\gamma\:\!\delta}	$ is the Weyl tensor, the $(s_A^\mu)_{A\in\{1,2\}}$ is the Sachs basis, and the expansion rate expressed in the form:
\begin{equation}\label{eqn:expansion2}
\theta=\frac{1}{d_A}\,\frac{\ud}{\ud\lambda}\,d_A\,.
\end{equation}

The procedure of finding the angular diameter distance $d_A$ is the following. For a given geodesic $x^\mu(\lambda)$, the first step is to fix the Sachs basis at the observer position $\lambda=0$. Sachs basis vectors $s_A^\mu$ are orthogonal to the the observer four-velocity, they are orthogonal to the direction of the incoming photon $d^\mu=(0,k^i)$, and they form the orthonormal basis on the two-dimensional screen, which reads:
\begin{equation}
s_A^\mu\,U_\mu=0\,,\quad s_A^\mu\,d_\mu=0\,,\quad s_A^\mu\,s_B^\nu\,g_{\mu\:\!\nu}=\delta_{A\:\!B}\,.
\end{equation}
These conditions leave some freedom, because one can choose the direction of one of these vectors on the two-dimesnional screen and the sign of the second vector arbitrary. We generate these arbitrary quantities randomly. After we construct the Sachs basis at the observer position, we solve numerically the parallel transport equations:
\begin{equation}
\frac{\ud}{\ud\lambda}\,s_A^\mu=-k^\rho\,\Gamma^\nu_{\rho\:\!\alpha}\,s_A^\alpha\,\quad A=1,2\,,
\end{equation}
by using the fourth order Runge-Kutta method. To test the numerical precision we verify the condition $s_A^\mu\,s_B^\nu\,g_{\mu\:\!\nu}=\delta_{A\:\!B}$ for each $\lambda>0$. This way we obtain the Sachs basis at each point on the geodesic $s_A^\mu(\lambda)$. After that, we solve with the same numerical method the system of equations (\ref{eqn:Focusing},\ref{eqn:Sachs_sigma}) with the initial conditions at the observer position $\lambda=0$ given by: $d_A=\epsilon\equiv10^{-6}$, $\ud d_A/\ud\lambda=1$, $\sigma_1=0$, $\sigma_2=0$. By taking the infinitesimally small initial value of the angular diameter distance $\epsilon\ll 1$ we avoid the singularity of the expansion rate given by (\ref{eqn:expansion2}) at $\lambda=0$.

By applying the procedure described above to the sample of geodesics generated in the Section \ref{sec:redshift} we found that the shear term $|\sigma|^2$, which appears in the focusing equation, is very small in comparison to the Ricci contribution $R_{\mu\:\!\nu}\,k^\mu\,k^\nu$. Putting $\sigma=0$ and solving numerically the focusing equation only provides the resulting angular diameter distance $d_A$, which is indistinguishable from the $d_A$ calculated from the full system of equations (\ref{eqn:Focusing},\ref{eqn:Sachs_sigma}). The difference at the redshift $z=1.2$ is $\Delta d_A\approx10^{-4}\,\mathrm{Mpc}$, and the relative difference is $\Delta d_A/d_A\approx 10^{-7}$. The \emph{no shear approximation} $\sigma\approx0$, which is sometimes considered in literature (e.g. \cite{1973ApJ...180L..31D}) is justified here. 

\begin{figure}[h]
   \centering
      \includegraphics[width=0.45\textwidth]{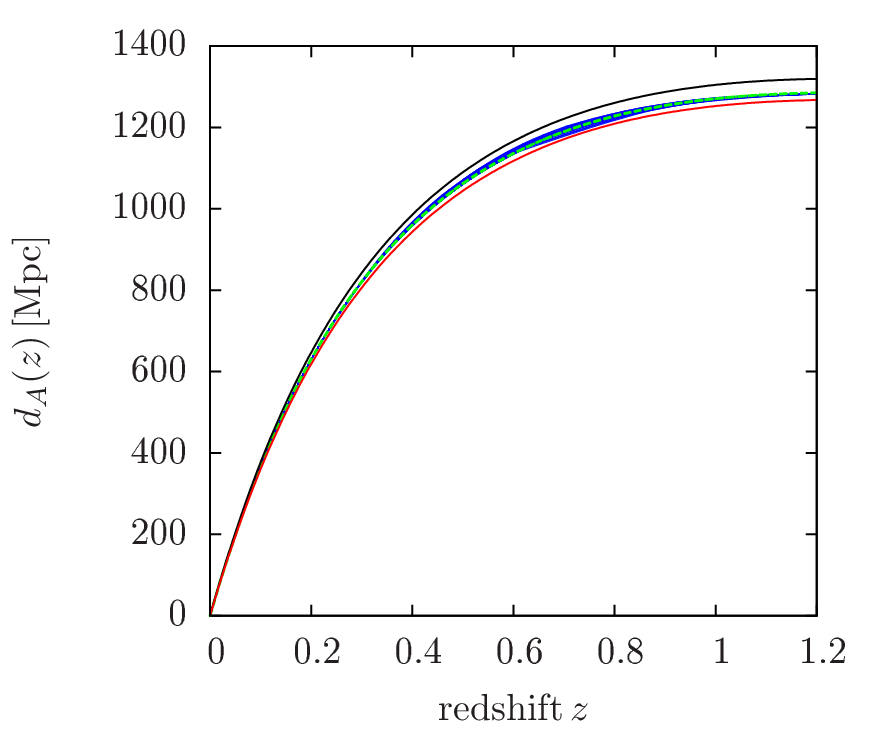}
      \caption{\label{fig:d_angular} \scriptsize{The angular diameter distance as a function of redshift calculated along each of the one hundred null geodesics of a random directions (\emph{the set of the blue curves}). This result is compared with the \mbox{Einstein}-\mbox{de Sitter} (\emph{black}), the FLRW model with parameters $\Omega_m=1.0497$, $\Omega_\Lambda=0.0$ (\emph{red}) and the Buchert average space-time (\emph{green dashes}).}}
\end{figure}
In Figure \ref{fig:d_angular} we present the angular diameter distance as a function of redshift $d_A(z)$, calculated with the numerical procedure described above for a sample of the one hundred geodesics generated in Section \ref{sec:redshift} (the blue curves). The origin of the spread of these results comes from the fact, that considered geodesics pass through different overdensity regions. For comparison, the reference relation $d_A(z)$ for the \mbox{Einstein}-\mbox{de Sitter} background is plotted by the black curve, while the $d_A(z)$ for the FLRW model with parameters $\Omega_m=\Omega^{(0)}+\langle \Omega^{(1)}\rangle_\mathcal{D}=1.0497$ and $\Omega_\Lambda=0.0$ is drawn by the red curve. On the same plot, the prediction of the $d_A(z)$ for a Buchert average space-time is given by the green dashed curve. By the Buchert average space-time we mean FLRW space-time with the line element $\ud s^2=-\ud t^2+a_\mathcal{D}^2(t)(\ud r^2+r^2\,\ud\Omega^2)$. To obtain the angular diameter distance in such a model we calculate numerically the integral:
\begin{equation}
d_A^{(Buchert)}(z)=\frac{1}{1+z}\,\int_{t_1}^{t_0}\frac{\ud t}{a_\mathcal{D}(t)}\,,
\end{equation}
where the $a_\mathcal{D}(t)$ is the effective scale factor calculated in Section \ref{sec:ModelParameters}, while the time of emission of light $t_1$ is given from the equation $a_\mathcal{D}(t_1)=1/(1+z)$.

\begin{figure}[h]
   \centering
      \includegraphics[width=0.45\textwidth]{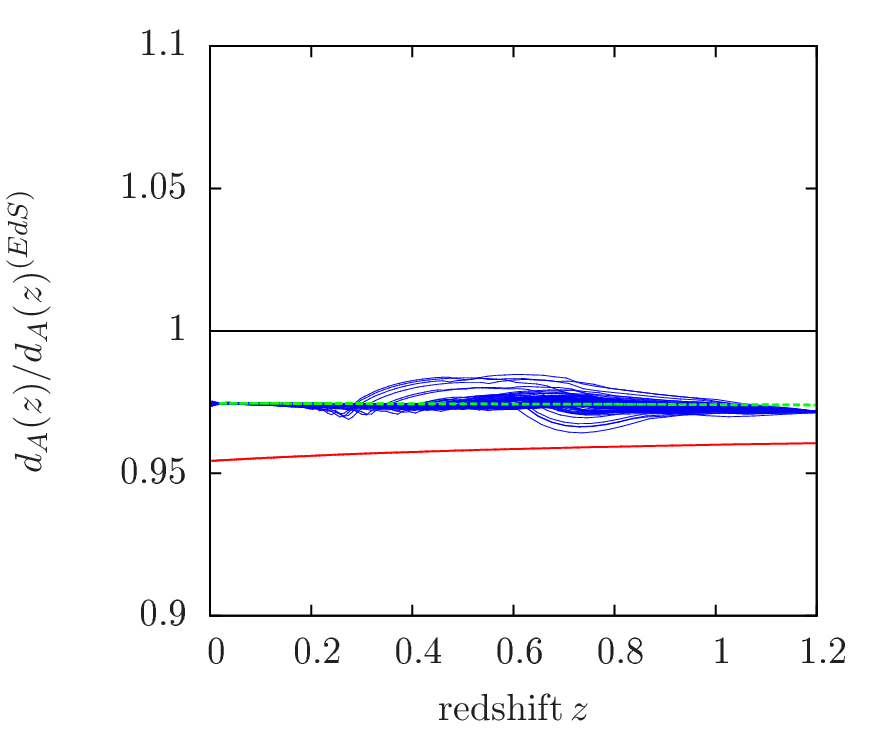}
      \caption{\label{fig:d_angular2} \scriptsize{The angular diameter distance relative to the \mbox{Einstein}-\mbox{de Sitter} model. The result of the numerical procedure described in the text (\emph{the set of the blue curves}) is compared with the FLRW model with parameters $\Omega_m=1.0497$, $\Omega_\Lambda=0.0$ (\emph{red}) and the Buchert average space-time (\emph{green dashes}).}}
\end{figure}
To show more clearly the differences between various models we present in Figure \ref{fig:d_angular2} the angular diameter distances relative to the \mbox{Einstein}-\mbox{de Sitter} one. It is seen that the angular diameter distances $d_A(z)$ calculated along the considered sample of geodesics (blue curves) differ significantly from the $d_A(z)$ predicted in the \mbox{Einstein}-\mbox{de Sitter} space-time (black line) and from $d_A(z)$ obtained in the FLRW dust model without cosmological constant, with the density equals to the average density of our inhomogeneous model (red curve). On the other hand, it seems that the $d_A(z)$ calculated in the Buchert average space-time fits well to the $d_A$-redshift relation derived numerically in the presented inhomogeneous model. It is of course not a proof that the Buchert approach is correct in a general situation, however, in this particular example, it gives the angular diameter distance as a function of redshift which is compatible with a direct numerical result.  

\section{Conclusions}
In this article, we presented explicitly the metric which approximately well describes the dust inhomogeneous cosmological model with periodically distributed overdensities. We obtained numerically the sample of one hundred random null geodesics, and for each of them we calculated the angular diameter distance as a function of redshift $d_A(z)$. By using the correspondence between the angular diameter distance and the luminosity distance \mbox{Eqn. \ref{eqn:luminosity}} one may translate these results onto luminosity distance-redshift relation $d_L(z)$, which can be derived from observations (e.g. by  supernovae Ia). In the considered situation the inhomogeneities have a nonnegligible impact on the $d_A(z)$.

Our example model can be instructive in comparison between various approaches to averaging in cosmology. From the viewpoint of the Green-Wald scheme, there is no backreaction effect in our space-time. However, according to \cite{2014CQGra..31w4003G}, \emph{the geodesics of the actual metric are not necessarily very close to corresponding geodesics of $g^{(0)}$}. Indeed, the $d_A(z)$ predicted in our model differs from the \mbox{Einstein}-\mbox{de Sitter} one. Therefore, the impact of the inhomogeneities onto the propagation of light should be taken into account separately, e.g. by the method described in the paper \cite{1998PhRvD..58f3501H}.  

From the viewpoint of the Buchert approach, it is interesting that the angular diameter distance obtained with the help of the effective scale factor fits to our results very well. This supports the statement that the effective scale factor $a_\mathcal{D}$ sometimes can be very useful, even if its nature is effective and not covariant. The similar conclusion concerning Tardis space-time one can find in \cite{2013JCAP...12..051L}. 

In a what sense the inhomogeneities could mimic accelerated expansion here? The angular diameter distance in our model with the average density equal to $1.0497$ in critical units is slightly higher than the angular diameter distance in the FLRW space-time with the same density. Then, we may expect that the fitting of the FLRW parameters to the resulting $d_A(z)$ profile causes a slight underestimation of the $\Omega_m$ parameter or a similar overestimation of the $\Omega_\Lambda$ parameter. 

The presented model space-time represents some simple toy model, which can be thought as a good starting point for further investigations.

\section*{Acknowledgement}
We would like to thank Sebastian Szybka, Boudewijn Roukema and Łukasz Bratek for their useful comments.

\bibliography{SikoraGlod2016}
\bibliographystyle{unsrt}

\end{document}